\documentclass{IEEEcsmag}

\usepackage[colorlinks,urlcolor=blue,linkcolor=blue,citecolor=blue]{hyperref}

\usepackage{upmath}
\usepackage{todonotes}
\usepackage{amssymb}
\usepackage{slashbox}

\usepackage{listings}
\usepackage{xcolor}

\definecolor{codegreen}{rgb}{0,0.6,0}
\definecolor{codegray}{rgb}{0.5,0.5,0.5}
\definecolor{codepurple}{rgb}{0.58,0,0.82}
\definecolor{backcolour}{rgb}{0.95,0.95,0.92}

\lstdefinestyle{mystyle}{
    backgroundcolor=\color{backcolour},   
    commentstyle=\color{codegreen},
    keywordstyle=\color{magenta},
    numberstyle=\tiny\color{codegray},
    stringstyle=\color{codepurple},
    basicstyle=\ttfamily\footnotesize,
    breakatwhitespace=false,         
    breaklines=true,                 
    captionpos=b,                    
    keepspaces=true,                 
    numbers=left,                    
    numbersep=5pt,                  
    showspaces=false,                
    showstringspaces=false,
    showtabs=false,                  
    tabsize=2
}

\newcommand*{\nolink}[1]{%
  \begin{NoHyper}#1\end{NoHyper}%
}

\usepackage[defaultcolor=red, addedmarkup=uuline]{changes}

\jvol{XX}
\jnum{XX}
\paper{8}
\jmonth{Dec}
\jname{Computer}
\pubyear{2022}

\begin{document}

\title{Overcoming the Barriers of Using Linked Open Data in Smart City Applications}

\author{Javier Conde}
\affil{Departamento de Ingeniería de Sistemas Telemáticos, Escuela Técnica Superior de Ingenieros de Telecomunicación, Universidad Politécnica de Madrid}
\author{Andres Munoz-Arcentales}
\affil{Departamento de Ingeniería de Sistemas Telemáticos, Escuela Técnica Superior de Ingenieros de Telecomunicación, Universidad Politécnica de Madrid}
\author{Johnny Choque}
\affil{Departamento de Ingeniería de Comunicaciones, Universidad de Cantabria}
\author{Gabriel Huecas}
\affil{Departamento de Ingeniería de Sistemas Telemáticos, Escuela Técnica Superior de Ingenieros de Telecomunicación, Universidad Politécnica de Madrid}
\author{Álvaro Alonso}
\affil{Departamento de Ingeniería de Sistemas Telemáticos, Escuela Técnica Superior de Ingenieros de Telecomunicación, Universidad Politécnica de Madrid}

\markboth{Department Head}{Overcoming the Barriers of Using Linked Open Data in Smart City Applications}

\begin{abstract}

Accurate, reliable, and highly available data is one of the critical aspects of the proper development of a smart city. In this article, we study the benefits and challenges of using Linked Open Data in smart city applications highlighting their main adoption barriers like the lack of standardization, and the difficulties of publishing and consuming Open Data sources. Our research proposes a set of open-source, high-scalable, and mature tools to overcome these challenges, providing a complete description of how those components can be integrated to enable and improve the way to publish and consume Linked Open Data in the ecosystem of smart city applications and its technological enablers (IoT, Artificial Intelligence, etc.).  Finally, we validate our proposal within the case of a public-rental bicycle system. The set of tools presented is agnostic to the field of application, thus, this case acts as a reference guide for other smart city applications.
\end{abstract}

\maketitle

\chapterinitial{Data are one of the most valuable} resources of the current society. The increasing demands of citizens have supported the rise of smart cities, i.e., cities that use technological infrastructure and data sources to improve the quality of life of their citizens \cite{future_trends_and_current_state_of_smart_city}. Smart cities carry out a deep digital transformation that affects all the environment, from small enterprises to big companies, governments, and citizens. To achieve this goal, high-quality data that are easy to access and share are required. With this aim, Beckwith et al. proposed Open Data (OD), i.e. accessible and free to use data, to build smart city applications \cite{data_flow_in_the_smart_city}.

In 2006, Tim Berners-Lee coined the term Linked Data (LD) as the enabling technology to manage the Web of Data (the Web where people and machines can access and navigate increasing their knowledge through connected data) \cite{linked_data_the_story}. He completed the concept of LD with the possibility of being realized under an open license, resulting in a new term known as Linked Open Data (LOD). With LOD, Tim Berners-Lee refers to sources of data that are public, updated, reusable, easy to share, and easy to access for people and machines. 

Among the actions to avoid corruption, governments started to publish their data openly to be transparent \cite{a_systematic_review_of_open_government}. However, according to Lnenicka and Nikiforova \cite{transparency_by_design_what_is}, transparency does not only mean to be public, it also implies to be understandable for the user (a person or a machine). That is, data must be served in an easy format to access, manage, and share. In this sense, Attard et al. \cite{a_systematic_review_of_open_government} make the distinction between public and OD, being OD, public data that are represented in a standardized format and accessible online. During the last decade, OD portals have appeared, defined by Ojo et al. \cite{realizing_the_innovation_potentials} as software tools that enable the interaction between users and OD through the publication, search, discovery, analysis, and visualization of data sets. According to Open Data Inception (\url{https://opendatainception.io}), nowadays there are almost 4000 portals distributed in more than 200 countries. Examples of the most large portals are the US Government OD portal, called data.gov (\url{https://www.data.gov/}); or the former European Data Portal, called today data.europa.eu (\url{https://data.europa.eu/}), which indexes datasets from other European portals. Not only governments publish OD, companies also do it. However, it is less frequent because companies refuse to freely share their data and to assume the costs related to the publication process \cite{open_data_hopes_and_fears}. The research conducted by Hurbean et al. \cite{open_data_based_machine_learning_applications} proposes the Internet of Things (IOT), and Machine Learning (ML) in combination with LOD platforms as the key elements for building smart city applications.

The aim of our work is to investigate the barriers present when using LOD in smart city applications, to propose a set of scalable open-source tools to overcome these challenges, and to validate the proposal through a use case.  

The article is structured as follows. The following section analyzes the use of LOD in smart city applications, focusing on the most common barriers from the perspective of data providers and data consumers. The next one describes a set of tools and good practices for building ML smart city applications using LOD. Then, we present a case of smart mobility in the city of Santander, applying the proposed tools. Finally, conclusions drawn from the article and future research are presented.

\section{OPEN DATA IN SMART CITY APPLICATIONS}

In the introduction, we mentioned that LOD supports the digital transformation of cities by providing a mechanism to publish a consume source of information in a free and understandable way \cite{data_flow_in_the_smart_city, digital_transformation_and_open_data, open_data_based_machine_learning_applications}. In the LOD ecosystem, there are two main stakeholders: data publishers (providers) and data users (consumers). Providers make data available to consumers who can access data directly for their analysis or building external services (e.g., smart city applications). In their research, Lnenicka and Komarkova \cite{big_and_open_linked_data_analytis_ecosystem} propose a finer division. They distinguish among (1) orchestrators who define the regulation and policies; (2) technology providers who manage the infrastructure; (3) producers who are the owners of the data; (4) publishers who process the data to make them accessible; (5) users who consume the data; and (6) prosumers who produce and consume data simultaneously. Regarding the publication and consumption of data, Attard et al. define the life cycle of open government data \cite{a_systematic_review_of_open_government}. The LOD cycle starts with new data created by the producer. The producer selects which data to publish, removing sensitive and private data. Subsequently, data publishers harmonize the information by making it accessible, understandable, and adding links to other sources. Finally, publishers publish the data and maintain them over time. Once data are open, consumers can discover them through OD portals, explore datasets, and exploit them for any specific purpose.

New data sources such as social media, social networks, or government data, and enabling technologies such as LOD, cloud computing, or IoT have allowed the evolution and growth of smart city applications \cite{future_trends_and_current_state_of_smart_city, big_data_applications_in_smart_cities}. However, new challenges have appeared. In the literature, authors analyze the barriers of using LOD for smart city applications in all stages of the LOD life cycle \cite{open_data_hopes_and_fears}. We are going to follow the approach of Attard et al. \cite{a_systematic_review_of_open_government} who divide challenges between providers and consumers.

\subsection{Provider barriers}

The purpose of providers is to make data available to consumers. Occasionally, they have to deal with a large amount of data that cannot be processed by a single machine. Providers also need to decide which data are going to be public and the license under they are going to be offered. Therefore, these raw data must be stored and preprocessed to make them transparent in the future. The existence of Big Data tools and techniques to automate these preprocessing tasks for the generation of LOD are required. Transparency does not only mean public, it also implies being accessible and understandable by machines and people \cite{transparency_by_design_what_is}. The annual report from data.europa.eu shows an improvement in this challenge. In 2015, only 35\% of the countries indexed on data.europa.eu provided at least 50\% of machine-readable datasets \cite{open_data_maturity_in_europe_2015}. In 2021 this percentage has increased to 93\% of the countries \cite{open_data_maturity_report_2021}. On the other hand, although data are machine-readable, the lack of data standardization difficulties the adoption of new data sources. For example, a smart city application obtains weather data from different sources that refer to the temperature parameter differently (e.g, temperature, temp), or using different units (e.g., K, ºC). This heterogeneity difficulties the development and maintenance of smart city applications, and as the amount of data sources increases the complexity increases too. The use of ontologies help to reduce this barrier, however tools for adapting the original data to the ontology are required. Another common issue is the lack of motivation of data providers who have to maintain the data and the platform that serves these data without perceiving a direct economic benefit from it \cite{open_data_hopes_and_fears}. OD portals act as an intermediary between consumers and providers, helping providers with the maintenance and publication process. OD portals have to deal with heterogeneous data sources, and as they aggregate more data sources, more difficulties appear, and a higher level of automation is required \cite{open_data_based_machine_learning_applications}. Considering that interoperability is one of the main pillars on which any LOD initiative should be built, the World Wide Web Consortium (\href{https://www.w3.org/}{https://www.w3.org/}) published the Data Catalog Vocabulary (DCAT [\url{https://www.w3.org/TR/vocab-dcat/}]) to facilitate the exchange of data catalogs between different organizations.

\subsection{Consumer barriers}

The aim of consumers is to exploit OD for their own benefit. For this purpose, they have access to a large number of platforms, so it is difficult for them to find where the desired datasets are located. In their work, Attard et al. \cite{a_systematic_review_of_open_government} distinguish three ways of publishing data. The first one are providers that publish their data on their web; the second one publish their data in OD portals; and the last one publish only the metadata in OD portals and these metadata point to the original source of data. The last alternative could be generalized to build OD portals that aggregate other portals in a process called harvesting \cite{a_systematic_review_of_open_government}. This approach helps consumers because they only have to consult the most generic data portals, which, in most cases, offer good search engines and SPARQL (\url{https://www.w3.org/TR/2013/REC-sparql11-overview-20130321/}) endpoints. Aggregator portals do not include data, they only store the metadata and point to the respective OD portals. As an example, data.europa.eu included 170 catalogs from 36 countries in January 2022.

Locating data sources is not the only barrier consumers have to deal with. They must also filter datasets based on the quality of the data. Common issues found are outdated data, incomplete data, inaccurate data, or data that do not fit the standards. Some portals such as data.europa.eu include mechanisms to assess the quality of data and metadata \cite{linked_data_in_the_european_data_portal}. In addition, consumers may have problems when accessing data due to the format they are served, for example, a dataset shared as a PDF is difficult to process by a machine \cite{transparency_by_design_what_is}. We have mentioned the potential of ML in combination with LOD to develop smart city applications. However, many data sources do not include historical data, which are fundamental for training ML algorithms. Big Data tools for consuming updates of the data, processing them, and building historical data help to reduce this issue. Other barriers are the legal ones, for example, new laws that limit some data sources or restrictive licenses \cite{open_data_hopes_and_fears}.

\begin{figure}
\centerline{\includegraphics[width=0.48\textwidth]{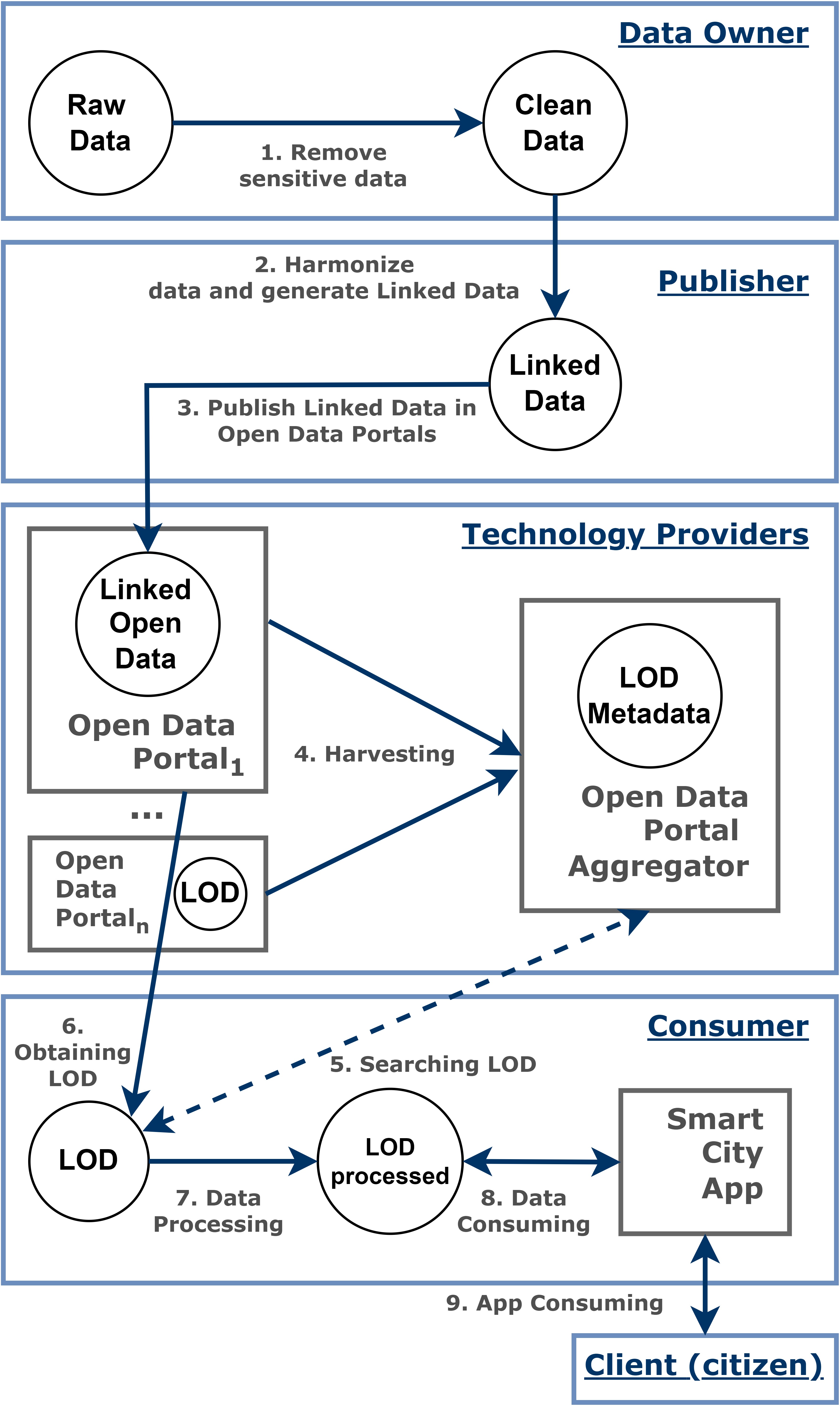}}
\caption{Usage of open data in smart city applications}
\label{fig:lod_smart_cities_arch}
\end{figure}

\nolink{\textbf{Figure~\ref{fig:lod_smart_cities_arch}}} represents the usage of LOD in smart city applications.

\section{TOOLS FOR OVERCOMING THE BARRIERS OF LINKED OPEN DATA IN SMART CITY APPLICATIONS}

There exist open source tools that can help overcome the barriers mentioned above. These tools are constantly evolving by adopting new technologies, which allows them to improve their functionalities. The drawback is that many of them focus on developing specific tasks, but they do not include components to integrate them and build complex smart city applications based on LOD. For this reason, in this section we present a set of tools that are easy to integrate with each other and with other systems, that are scalable and compatible with Big Data scenarios, that have an active community, and that are continuously evolving to adapt to new necessities. They are mature technologies, tested in numerous use cases, as the smart mobility case presented in this work. We also highlight those tools that we have implemented to fill gaps that have been identified during the development of smart city applications.

Some of the proposed tools belong to the \textbf{FIWARE} ecosystem (\href{https://www.fiware.org/}{https://www.fiware.org/}). FIWARE is a framework of open-source components, known as Generic Enablers (GEs [\url{https://github.com/FIWARE/catalogue}]), that facilitate the development and implementation of smart solutions and management of context information, including smart city applications and Digital Twins \cite{Modeling_Digital_Twin_Data_and_Architecture}. The core component of FIWARE is the \textbf{Orion-LD Context Broker GE}, (\url{https://github.com/FIWARE/context.Orion-LD}). The Context Broker manages context information through the implementation of a publish-subscribe pattern providing a Next Generation Service Interface-Linked Data (NGSI-LD \cite{ngsi_ld_specification}), an evolution of the NGSIv2 interface (\url{https://fiware.github.io/specifications/ngsiv2/stable/}) for managing context information following the principles of LD. Clients communicate with the Context Broker through HTTP/HTTPs, accessing directly the context information, or subscribing to changes in order to receive notifications when they happen. Other components of the FIWARE ecosystem are \textbf{Draco} (\url{https://github.com/ging/fiware-draco}) to transform and persist data; \textbf{Iot Agents} (\url{https://github.com/FIWARE/catalogue/tree/master/iot-agents}) to connect IoT devices; \textbf{Cosmos} (\url{https://github.com/ging/fiware-cosmos-orion-spark-connector}) to enable Big Data analysis; etc.

The centerpiece of our proposal for publishing metadata is \textbf{CKAN} (\url{https://github.com/ckan}). CKAN is an open-source software which main goal is to provide a managed data-catalog-system for OD. It helps providers publish their datasets and make their data transparent to the public. In fact, data.europa.eu promotes the use of CKAN in OD portals that wish to be indexed (i.e., harvested) on its platform.
 
CKAN facilitates the publication of datasets, but it does not solve the lack of standardization in data and metadata. Regarding data modeling, we propose NGSI-LD. NGSI-LD is agnostic to its domain of application, and it was standardized by ETSI (European Telecommunications Standards Institute [\url{https://www.etsi.org/}]). The API defines a set of HTTP methods to create, read, update, and delete entities \cite{enabling_context_aware_data_analytics}. Thus, with the adoption of the NGSI standard, a new initiative called Smart Data Models (\url{https://github.com/smart-data-models}) has emerged. The main goal of the Smart Data models is to integrate reference data models, approved by the industry, with the NGSI-LD API, adding to the usual mechanism of accessing the context information, the possibility to standardize the data. Regarding metadata, we propose using \textbf{DCAT}, a specification defined in RDF (\url{https://www.w3.org/TR/rdf11-primer/}) that allows the description of datasets through curated collections of metadata. In addition, DCAT provides the possibility to define profiles, which are a customized specification that adds additional constraints. One of them is the DCAT Application Profile for Data Portals in Europe (\textbf{DCAT-AP} \cite{dcat_ap_specification}). Standardizing metadata through DCAT and its profiles ensures semantic interoperability between different OD portals, easing the harvesting process, and helping consumers (people or machines) find datasets using metadata as filter criteria \cite{linked_data_in_the_european_data_portal}. The compatibility of DCAT, and its profiles like DCAT-AP, with CKAN portals must be granted. The \textbf{DCAT CKAN extension} (\url{https://github.com/ckan/ckanext-dcat}) maps the metadata in CKAN format to the metadata in DCAT format. For example, CKAN tags are assigned DCAT keywords. We have improved this extension creating a new one called \textbf{CKAN DCAT-AP} (\url{https://github.com/YourOpenDAta/ckanext-dcatapedp}) that implements the DCAT-AP profile. 

Data and metadata standardization is granted with DCAT (DCAT-AP), NGSI-LD, and the Smart Data Models. However, it is common for providers to work with heterogeneous data sources that have to be preprocessed. For these preprocesing tasks, we propose the \textbf{Draco GE}. Draco is a dataflow management system based on Apache NiFi that is highly scalable and suitable for data transforming operations and data routing through a set of highly configurable processors and controllers. Draco eases the ingestion of data from different sources using different protocols like HTTP (\texttt{InvokeHTTP} and \texttt{ListenHTTP} processors); transforming the data into some standards like NGSI-LD (\texttt{JoltTransfromJson} processor); and generating metadata to be published on CKAN portals. In order to implement the latest functionality, we have developed the \texttt{UpdateCKANMetadata} and \texttt{NGSIToCKAN} processors. The \textbf{UpdateCKANMetadata} processor creates metadata that are compliant with CKAN. This issue is not easy to solve because data sources are very heterogeneous and data can be retrieved in different formats, languages, etc. This processor is intended to be used jointly with the \textbf{NGSIToCKAN} processor to persist NGSI-LD context data and metadata within a CKAN server, creating all the structures required (organizations, packages, resources, etc.). Moreover, Draco is needed to maintain the historical evolution of the data. Every time a change in any entity occurs, Draco takes care of storing the updated entity values in one of the supported database management systems (PostgreSQL, MongoDB, MySQL, etc.). It is important to remark that these recorded data are extremely valuable for ML models due to the fact that these data are used in the training stage of such systems. Besides Draco, we propose the \textbf{IoT Agents} to integrate the data served by IoT sensors in the Context Broker. The Iot Agent models the information as NGSI Smart Data Model entities. IoT Agents manage the interaction between the North NGSI and the South IoT Specific protocols like JSON, UltraLight over HTTP/MQTT, or LoRaWAN. These components abstract the communication protocols used by IoT devices and translate them into the FIWARE NGSI format.

Once data are offered as open, consumers need filter capabilities to find them. CKAN allows to filter datasets through their metadata. \textbf{SPARQL} queries can be used for more specific filters, but metadata in DCAT format must be stored in a triple store. An important criterion for filtering datasets is the quality of data and metadata. In this case, we propose \textbf{Metadata Quality Assessment (MQA) Tools}, i.e., pieces of software that assess the quality of datasets. For example, data.europa.eu defines a set of indicators to evaluate the catalogs harvested \cite{linked_data_in_the_european_data_portal}. Each indicator must meet a set of conditions based on the use of one or more properties of the DCAT-AP vocabulary. The MQA methodology defines a weight for each indicator according to its importance. Finally, the overall scoring of the dataset is calculated and the dataset is labeled as excellent, good, sufficient, or bad. In January 2022, 1,397,750 datasets were indexed in data.europa.eu. Of all of them, only 124 (0.009\%) were labeled as excellent; 233,867 (16.731\%) as good; 677,400 (48.464\%) as sufficient; and 486,359 (34.796\%) as bad. The Eruopean portal provides a tool to validate DCAT-AP files using Shapes Constraint Language \textbf{(SHACL)} specifications (\url{https://data.europa.eu/mqa/shacl-validator-ui/}). This tool tests the vocabulary and syntax of the metadata included in the dataset but does not calculate the total score. To cover this gap, we have developed the tool \textbf{mqa-scoring-api} (\url{https://github.com/YourOpenDAta/mqa-scoring-api}) which verify that the requirements specified by the MQA are met for each indicator included in the dataset. With this tool, providers can improve the quality of their datasets before being harvested by aggregator portals.

Regarding processing data, consumers sometimes have to deal with huge datasets. The \textbf{Cosmos GE} provides Big Data capabilities in the FIWARE ecosystem. Cosmos is a set of connectors that allow data ingesting from the Context Broker directly within two of the most popular open-source data processing frameworks for Big Data: Apache Flink (\url{https://github.com/apache/flink}) and Apache Spark (\url{https://github.com/apache/spark}).

\nolink{\textbf{Table~\ref{table:cycle_roles}}} summarizes the set of tools proposed. It shows in which phase of the OD life cycle they are used and by which stakeholder.

\begin{table*}
\caption{Roles, tools, and operations involved in the Open Data life cycle.}
\label{table:cycle_roles}
\footnotesize
\begin{tabular*}{36pc}{|p{81pt}<{\raggedright}|p{65pt}<{\raggedright}|p{75pt}<{\raggedright}|p{80pt}<{\raggedright}|p{69pt}<{\raggedright}|@{}}
\cline{2-4}
\multicolumn{1}{l|}{} & \multicolumn{3}{c|}{Data provider} & \multicolumn{1}{l}{} \\ \cline{2-5} 
\multicolumn{1}{l|}{} & Data Producer & Data Publisher & Technology Provider & Data Consumer \\ 
\hline
Creation \& Selection & - IoT sensors, web servers, third-part systems. & - IoT Agent{\scriptsize $^{1}$}, Draco{\scriptsize$^{2}$}.  &  &  \\
Harmonization &  & - NGSI-LD{\scriptsize$^{3}$}, Smart Data Models{\scriptsize$^{4}$}. & - Orion-LD Context Broker{\scriptsize$^{5}$}. &  \\
Publication \& Linking & & - Draco{\scriptsize$^{2}$}. & - CKAN server{\scriptsize$^{6}$},  RDF{\scriptsize$^{7}$} {\scriptsize(DCAT$^{8}$/DCAT-AP$^{9}$)}. &  \\
Curation &  &  & - SHACL shapes{\scriptsize$^{10}$}, MQA Tools{\scriptsize$^{11}$}. &  \\ 
Discovery \& Exploration &  &  & - Harvesters, SPARQL engine{\scriptsize$^{12}$}, CKAN search engine {\scriptsize$^{6}$} & - Data analysis tools. \\ 
Exploitation &  &  & - APIs. & - Cosmos{\scriptsize$^{13}$}, Draco{\scriptsize$^{2}$}, web servers, AI, visualization tools, databases, etc. \\  
\hline
\multicolumn{5}{@{}p{36pc}@{}}{\scriptsize
{\tiny$^{1}$}IoT Agents GE: \url{https://github.com/FIWARE/catalogue/tree/master/iot-agents},
{\tiny$^{2}$}Draco GE: \url{https://github.com/ging/fiware-draco}, 
{\tiny$^{3}$}NGSI-LD: \cite{ngsi_ld_specification},
{\tiny$^{4}$}Smart Data Models: \url{https://github.com/smart-data-models},
{\tiny$^{5}$}Orion-LD GE: \url{https://github.com/FIWARE/context.Orion-LD},
{\tiny$^{6}$}CKAN: \url{https://github.com/ckan},
{\tiny$^{7}$}RDF: \url{https://www.w3.org/TR/rdf11-primer/},
{\tiny$^{8}$}DCAT: \url{https://www.w3.org/TR/vocab-dcat/},
{\tiny$^{9}$}DCAT-AP: \cite{dcat_ap_specification},
{\tiny$^{10}$}SHACL validator: \url{https://data.europa.eu/mqa/shacl-validator-ui/},
{\tiny$^{11}$}MQA Scoring Api Tool: \url{https://github.com/YourOpenDAta/mqa-scoring-api},
{\tiny$^{12}$}SPARQL: \url{https://www.w3.org/TR/2013/REC-sparql11-overview-20130321/},
{\tiny$^{13}$}Cosmos GE: \url{https://github.com/ging/fiware-cosmos-orion-spark-connector}.
}\\
\end{tabular*}
\end{table*}

\section{USE CASE}

\begin{figure*}
\centerline{\includegraphics[width=30pc]{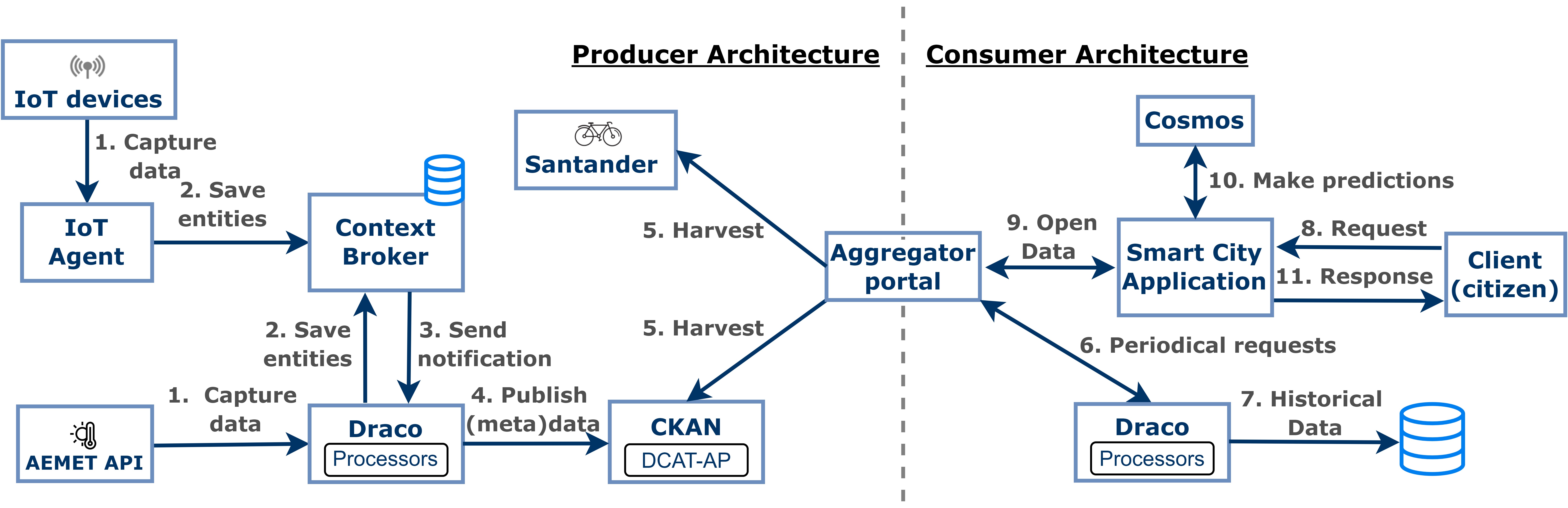}}
\caption{Provider and consumer bicycles rental service architecture}
\label{fig:architecture}
\end{figure*}

In this section, we present a use case of smart mobility to validate the set of tools proposed to overcome the barriers of OD in smart city applications. In the research conducted by Kirimtat et al. \cite{future_trends_and_current_state_of_smart_city} they detected smart mobility as one of the most relevant topics in smart cities. This use case shows an application that predicts the availability of rental bikes parked at public bike stations in the city of Santander, Spain. \nolink{\textbf{Figure~\ref{fig:architecture}}} shows two parallel architectures, one focused on data providers and the other on data consumers.

\subsection{Providers architecture}
Providers generate data and publish them transparently to consumers. In the bicycle rental case, we distinguish among three data providers: the Santander City Council, the meteorological Spanish agency called AEMET, and IoT sensors spread throughout the city of Santander. Santander City Council publishes in real time the number of bicycles parked at public stations. The information is included in the Santander OD portal (\href{http://datos.santander.es/}{http://datos.santander.es/}), which is harvested by data.europa.eu. AEMET (\href{http://www.aemet.es/}{http://www.aemet.es/}) provides data from meteorological stations. This information is offered by a REST API but is not included in any OD portal. IoT sensors are located on the street, they capture traffic intensity, and they are not published in any OD portal.

Consequently, the first task of providers is to publish the AEMET data on OD portals to make them transparent and harvestable by aggregator portals. For this purpose, Draco is configured with a set of processors. Draco periodically requests to AEMET for the last weather information of Santander using an \texttt{InvokeHTTP} processor. It receives the following data (summarized for the sake of simplicity): 
\begin{lstlisting}
{
  "ubi" : "SANTANDER",
  "fint" : "2021-11-10T15:00:00",
  "prec" : 0.0,
  "ta" : 14.6
}
\end{lstlisting}

Next, Draco transforms the incoming data into a \texttt{WeatherObserved} entity using a \texttt{JoltTransformJSON} processor. The resultant entity is both compliant with the NGSI-LD standard and with the Smart Data Models. Finally, Draco publishes the entity to the FIWARE Context Broker using another \texttt{InvokeHTTP} processor: 
\begin{lstlisting}
{
  "id": "urn:WeatherObserved:Santander",
  "type": "WeatherObserved",
  "address": {
    "addressLocality": "Santander",
    "addressCountry": "ES"
  },
  "dateObserved":"2021-11-10T15:00:00.00Z",
  "precipitation": 0,
  "temperature": 14.6,
  "@context":["https://smartdatamdels.org/context.jsonld"]
}
\end{lstlisting}

In parallel, Draco receives a notification when a new entity appears in the Context Broker. It generates the metadata using the \texttt{UpdateCKANMetadata} processor and creates the dataset in the CKAN portal with the \texttt{NGSIToCKAN} processor. Finally, the CKAN extension adapts the metadata to be compliant with the DCAT-AP standard, the one recommended by data.europa.eu for being harvested:

\begin{lstlisting}
<?xml version="1.0" encoding="utf-8"?>
<rdf:RDF 
 xmlns:rdf="http://www.w3.org/1999/02/22-rdf-syntax-ns#" 
 xmlns:dcat="http://www.w3.org/ns/dcat#" 
 xmlns:dct="http://purl.org/dc/terms/">
 <dcat:Dataset>
  <dct:title>
   Santander AEMET Weather
  </dct:title>
  <dct:description>
   Santander weather in real time
  </dct:description>
  <dcat:distribution>
   <dcat:Distribution>
    <dct:title>
     Santander WeatherObserved Entity
    </dct:title>
    ...
   </dcat:Distribution>
  </dcat:distribution>
 ...
 </dcat:Dataset>
</rdf:RDF>
\end{lstlisting}

Thanks to Draco's Big Data capabilities, the solution can be scaled to feed the CKAN portal with all the AEMET data, including all the cities of Spain and not only Santander.

Regarding IoT sensors, the process is similar. In this case, the FIWARE IoT Agent receives sensor measures through the MQTT protocol and acts as a bridge by transforming and saving the incoming data in the Context Broker modeled as NGSI-LD \texttt{TrafficFlowObserved} Smart Data Model entities. Then, the process is similar to the previous one: Draco uses the \texttt{UpdateCKANMetadata} and \texttt{NGSIToCKAN} processors to generate the metadata and the dataset in CKAN.

Data providers can check the quality of their data and metadata using MQA tools and fine-tune the processors to increase the quality of their datasets.

\subsection{Consumers architecture}
For the bicycle rental service case, data consumers are developers of smart city applications. First, they need to find the data sources. As mentioned above, providers publish their data on OD portals, generate metadata compliant with the DCAT-AP standard, and are harvested by aggregator portals. Harvesting makes it easier for consumers to find the datasets since they only explore the aggregator portal which acts as a single source of truth. Moreover, they can use SPARQL engines to filter the datasets through the metadata:
\begin{lstlisting}
PREFIX dcat: <http://www.w3.org/ns/dcat#>
PREFIX dct: <http://purl.org/dc/terms/>
PREFIX foaf: <http://xmlns.com/foaf/0.1/>
SELECT DISTINCT ?dataset  WHERE { ?dataset a dcat:Dataset . 
  ?dataset dct:publisher ?publisher .
  ?publisher foaf:name "Ayuntamiento de Santander" .
  ?dataset dct:description ?title .FILTER regex(str(?title), "bicycle") .
} LIMIT 10
\end{lstlisting}

The smart city application will allow citizens to predict the availability of bicycles in the future. For this purpose, supervised ML algorithms will be applied in Cosmos. ML models are trained using historical data from bicycle stations in combination with weather and traffic information. However, there does not exist any register of historical data. Consequently, Draco is going to be in charge of making periodic requests to these data sources and persisting the information in a historical database. A server will respond to citizen requests acting as the entry point of the system. 

The resultant flow of data starts when a citizen asks for a prediction and the server processes the request. First, it collects the last measure of bike stations, weather, and traffic from OD portals. Then, the server sends the request to Cosmos to make the prediction. Finally, the server answers the citizen with the result obtained.   

The architecture presented in \nolink{Figure~\ref{fig:architecture}} is oriented to Big Data environments, so it can be adapted to client demands. Moreover, the adoption of standards like NGSI-LD or Smart Data Models makes the solution easy to extend to other cities. Otherwise, it would be necessary to design, develop, and implement a new application from scratch.

As lessons learned from this use case, we highlight that the tools presented have been essential during the publication and integration of LOD, facilitating the development of the application, and solving the barriers of using LOD in smart city applications. As the technology presented is agnostic to the field of application, it can be extended to other scenarios, resulting this work as a reference guide to implement new solutions. Thanks to Draco and the IoT Agent, we have merged different data sources homogenizing them through the FIWARE Smart Data models, and making them accessible for machines through the Fiware Context Broker. Moreover, Draco has enabled the publication of the homogenized data in a CKAN OD portal, generating the metadata of the datasets compliant with the DCAT-AP profile. Once the data were published as OD, the consumer was able to process them using Cosmos for the ML capabilities. However, the lack of historical data has made it difficult to obtain better predictions of bicycle availability. We have used Draco to generate the historical data, but these data are not offered as Open to the rest of consumers.

\section{CONCLUSION}

The fast growth of cities and the increasing demands of their citizens led to the appearance of smart cities, making it necessary to define and standardize the whole process, being LOD, IoT, and AI the key elements for this transition. In this article, we presented the most common barriers in the LOD life cycle for both the perspectives of data providers and data consumers. We proposed NGSI-LD and the Smart Data Models initiative to standardize the data, and DCAT to standardize the metadata. We also presented a set of scalable open-source tools compliant with the standards that allow to overcome the challenges. On the one hand, the FIWARE GEs manage the data in all stages of the OD life cycle. On the other hand, CKAN and its extensions are proposed as the reference open-source tools for developing OD portals. CKAN fulfills all the requirements of OD and includes the mechanisms to be harvested by aggregator portals. We validated our proposal through a smart city rental bicycle application use case. We concluded that DCAT, NGSI-LD, the Smart Data Models, and all the tools presented are integrated in an easy and efficient way for both data consumers and providers. As future work, the list of tools could be extended and the proposal validated in different domains, exploring new smart city environments, and studying new architectures for consumers and providers.

\section{ACKNOWLEDGMENT}
This research was funded by Programa Propio UPM and co-financed by the Connection Europe Facility of the European Union under 2019-ES-IA-0121 of the
European Commission: Your Open DAta.

\bibliographystyle{IEEEtran}
\bibliography{IEEEabrv, ref}

\begin {IEEEbiography} {Javier Conde}{\,} is currently working as a researcher at UPM and pursuing a PhD in Telematics Engineering at the same university. His research interests include Big Data, Digital Twins, Linked Open Data, and Machine Learning.
\end{IEEEbiography}

\begin {IEEEbiography} {Andres Munoz-Arcentales}{\,} is currently working as researcher in the Department of Telematics Engineering at UPM. He obtained his Ph.D. in 2021 and in Telematics Engineering at the same university and his main research interests are Microservices, Data Fusion, Machine Learning, and Big Data.
\end{IEEEbiography}

\begin {IEEEbiography} {Johnny Choque}{\,} is currently working as a researcher in the Universidad de Cantabria where he received his Ph.D. in 2014. He is active on the IoT-enabled Smart Cities, Low Power Wireless Technologies, Open Data, and Blockchain.
\end{IEEEbiography}

\begin {IEEEbiography} {Gabriel Huecas}{\,} is currently working as an Associate Professor with the UPM. His main research interests are Big Data, Digitization, and Cloud Computing.
\end{IEEEbiography}

\begin {IEEEbiography} {Álvaro Alonso}{\,} is currently an Associate Professor with the UPM. His main research interests are Public Open Data, Security Management in Smart Context environments, and Multivideoconferencing Systems.
\end{IEEEbiography}

\end{document}